\documentclass[preprint]{emulateapj}   

\usepackage{mathptmx}

\journalinfo{The Astronomical Journal}
\submitted{Accepted by the Astronomical Journal}
\shorttitle{New Ultraviolet Observations of AM CVn}
\shortauthors{Wade, Eracleous, \& Flohic}

\begin{document}

\title{New Ultraviolet Observations of AM CVn
\footnote{B\lowercase{ased on observations made with the NASA/ESA
Hubble Space Telescope, obtained at the Space Telescope Science
Institute, which is operated by the Association of Universities for
Research in Astronomy, Inc., under NASA contract NAS 5-26555.}}}

\author{Richard A.\ Wade, Michael Eracleous\footnote{The Center for
Gravitational Wave Physics, The Pennsylvania State University,
Pennsylvania State University, University Park, PA 16802}, and
H\'el\`ene M.\ L.\ G.\ Flohic} \affil{Department of Astronomy \&
Astrophysics, The Pennsylvania State University,\\ University Park, PA
16802} \email{wade@astro.psu.edu, mce@astro.psu.edu,
flohic@astro.psu.edu}

\begin{abstract}
We have obtained observations of the ultraviolet spectrum of AM CVn,
an ultra-short-period helium cataclysmic variable, using the Space
Telescope Imaging Spectrograph (STIS) aboard the Hubble Space
Telescope (HST).  We obtained data in time-tag mode during two
consecutive orbits of HST, covering 1600--3150~\AA\ and
1140--1710~\AA, respectively.  The mean spectrum is approximately flat
in $f_\nu$.  The absorption profiles of the strong lines of
\ion{N}{5}, \ion{Si}{4}, \ion{C}{4}, \ion{He}{2}, and \ion{N}{4} are
blue-shifted and in some cases asymmetric, evidencing a wind that is
partly occulted by the accretion disk.  There is weak red-shifted
emission from \ion{N}{5} and \ion{He}{2}.  The profiles of these lines
vary mildly with time.  The light curve shows a decline of $\sim$20\%
over the span of the observations.  There is also flickering and a 27
s (or 54 s) ``dwarf nova oscillation'', revealed in a power-spectrum
analysis.  The amplitude of this oscillation is larger at shorter
wavelengths.  We assemble and illustrate the spectral energy
distribution (s.e.d.)  of AM CVn from the ultraviolet to the
near-infrared.  Modeling the accretion phenomenon in this binary
system can in principle lead to a robust estimate of the mass
accretion rate on to the central white dwarf, which is of great
interest in characterizing the evolutionary history of the binary
system.  Inferences about the mass accretion rate depend strongly on
the local radiative properties of the disk, as we illustrate.
Uncertainty in the distance of AM CVn and other parameters of the
binary system presently limit the ability to confidently infer the
mass accretion rate.
\end{abstract}



\section{Introduction}\label{sect:intro}

AM CVn (HZ 29; $\alpha_{2000}= 12^h 34^m 54\fs 6$,
$\delta_{2000}= +37^\circ 37\arcmin 43\arcsec$,
$l=140^\circ$, $b=+79^\circ$, $V\approx$14.1) is the type star of the
small class of helium cataclysmic variables (HeCVs). These are also
called interacting binary white dwarfs, although the mass donor star
which fills its Roche lobe and transfers mass via a gas stream and
accretion disk to the mass-gaining white dwarf may not itself be fully
degenerate. Hydrogen lines are absent from the spectra of these
objects. A recent review of the class and a summary of possible
evolutionary pathways leading to the AM CVn stars can be found in
Nelemans (2005).

The orbital period of AM CVn is $P_{\rm orb} = 1029$~s (Nelemans et
al.\ 2001).  A signal is found at this period in the power spectrum of
time-series photometry, and a second signal is found near $P_{\rm sh}
= 1051$~s. This latter signal is thought to represent a ``permanent
superhump'', at the beat period between the orbital period and the
precession period of a non-circular disk.  From the fractional excess
of $P_{\rm sh}$ over $P_{\rm orb}$, it is thought possible to estimate
the mass ratio of the binary system, $q= M_2/M_1$, where $M_2$ is the
mass of the donor star (see \S\ref{sect:obsdsed}).  A recent paper
describes a kinematic measurement of $q$ (Roelofs et al.\ 2006).

The accretion disk in AM CVn appears always to be in a stable ``high''
state, consistent with the relatively high mass transfer rate that is
expected of the HeCVs at the short end of the observed range of
periods.  (Smak 1983; Tsugawa \& Osaki 1997; Deloye et al.\ 2005).
Mass transfer is thought to be driven ultimately by gravitational wave
radiation of orbital angular momentum, combined with the expansion of
the donor star as mass is lost. In this model, the orbital period
evolves to larger values with time, and the mass transfer rate to
smaller values.  A {\em measurement} of the mass accretion rate can
help constrain the mass ratio and/or the donor star mass, from which
it may be possible to reconstruct the prior history of the binary
(Deloye et al.\ 2005).  Inferring the mass transfer rate from
observation involves measuring the distance of the system and
estimating the bolometric flux, or accurately modeling the spectral
energy distribution (s.e.d.) of the accretion disk, also taking into
account light from the two stars.  Either method involves measuring
the emitted spectrum over a broad range of wavelengths.

The visible spectrum of AM CVn has been studied extensively, and the
visible s.e.d.\ has been modeled on the assumption that it is
dominated by the accretion disk (e.g., El-Khoury \& Wickramasinghe
2000; Nasser et al.\ 2001; Nagel et al.\ 2004).  A reasonable degree
of success has been achieved in these studies in accounting for the
profiles of the \ion{He}{1} absorption lines, although different
authors have inferred (or imposed!)  different properties of the disk
such as the mass transfer rate, the disk size, and the disk
inclination. The ultraviolet (UV) spectrum and the overall s.e.d.\
have received less attention in recent years but offer the
possibility, when used together with the \ion{He}{1} lines, of more
powerfully constraining the parameters of the disk model.

With the above considerations in mind, we obtained observations of the
UV spectrum of AM CVn using the Space Telescope Imaging Spectrogram
(STIS) aboard the Hubble Space Telescope (HST).  One goal was
to model the UV absorption line spectrum of the accretion disk in some
detail and thus infer the mass transfer rate $dM/dt$ and inclination
angle $i$ of the disk rather directly.  We also hoped to learn more
about the chemical composition of the material being transferred.
Instead of showing an absorption spectrum that arises purely from a
steady disk in local hydrostatic equilibrium, however, the STIS
observations show evidence of a disk wind, which obliterates many of
the stronger diagnostic lines and blends that are key to such an
analysis.  Moreover, the existence of a wind calls into question some
of the standard assumptions underlying models of accretion disk
atmospheres.  

The STIS data greatly exceed, in quality, reliability, spectral
resolution, and time resolution, most of the UV observations of AM CVn
that were made with the International Ultraviolet Explorer (IUE) in
the period 1978-1991 (which nevertheless hinted at the presence of
wind line profiles).  The STIS data exceed in wavelength coverage some
earlier HST observations of AM CVn, made using the Goddard High
Resolution Spectrograph (GHRS). They have revealed some behavior not
previously reported, including oscillations in light level with a
period of $\sim$27 s.

We defer a detailed modeling effort and in this contribution focus on
describing the UV spectrum and its behavior and on presenting the
overall s.e.d.\ of AM CVn.  We describe the STIS observations and the
major features of the spectrum, emphasizing the wind line profiles,
their variation with time, and the time variation and power spectrum
of the measured fluxes.  We assemble the UV-visible-infrared s.e.d.\
of AM CVn, which forms one basis for modeling the accretion process in
this binary.  We briefly discuss some practicalities and uncertainties
of such modeling, which lead to uncertainties in the inferred mass
accretion rate.  In particular, we illustrate the critical role played
by the radiative properties of the gas in the accretion disk.  We also
emphasize the importance of an accurate knowledge of the binary mass
ratio and distance to AM CVn, as they relate to inferring $dM/dt$.

The paper is organized as follows. In \S 2 we describe the
observations with STIS, then present our findings concerning mean and
time-dependent behavior of the flux level and line profiles.  In \S 3
we discuss the apparent variation in UV flux level, identify the
{27-s} oscillations as ``dwarf nova oscillations'', compare the
``wind'' line profiles with theoretical expectations and observations
of other systems, and present and discuss the s.e.d.\ of AM CVn.  We
briefly summarize our findings and discussion in \S 4.


\section{Observations and Findings}\label{sect:obsns}

We observed AM CVn with STIS on 2002 February 21, during two
consecutive orbits of the HST.  The target was observed in time-tag
mode, in the first orbit with the NUV-MAMA and grating G230L for 32
minutes, and in the second orbit with the FUV-MAMA and grating G140L
for 40 minutes.\footnote{See http://www.stsci.edu/hst/stis/ for the
STIS Instrument Handbook.}  Thus we have complete orbital coverage of
the binary star with each detector/disperser setup, but we do not have
coverage of the 13.4~h precessional period of the eccentric disk which
gives rise to a 1051~s superhump period (Patterson et al.\ 1993).

The NUV observation started at 09:16:57 UT and lasted for 1920~s, with
wavelength coverage 1577--3161 \AA.  The FUV observation started at
10:35:48 UT and lasted for 2400~s, thus ending almost 2 hours after
the start of the NUV exposure. The FUV wavelength coverage was
1121--1717 \AA, overlapping the NUV coverage.  The ``52X0.1'' aperture
($0\farcs1$ wide, long slit) was used.  An Acquisition/Peak-up sequence of
exposures was done prior to the NUV observations to center the star
accurately in the slit. During the spectrum exposures, the
r.m.s. jitter of HST on the V2 and V3 axes was less than about 8.5
milli-arcseconds, with no recenterings and no losses of lock.

\subsection{Trend of Flux Level with Time}\label{sect:trend}

\begin{figure}
\includegraphics[angle=-90,scale=.37]{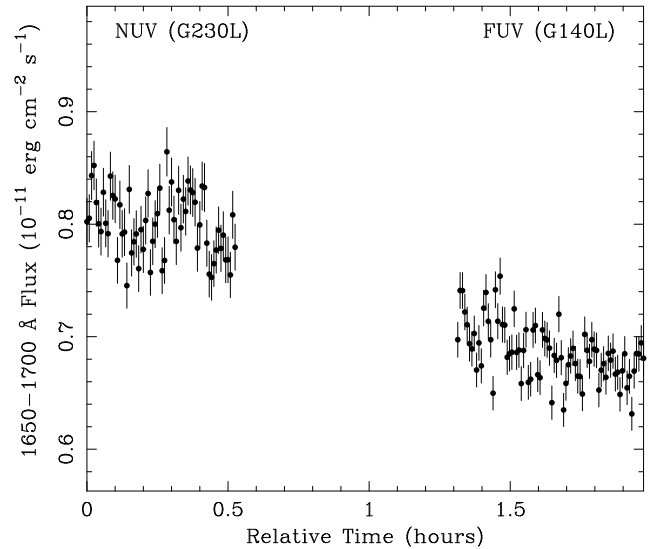}
\caption{Light curve of AM CVn showing the 1650--1700 \AA\ integrated
flux during the {\it HST} observation. Each point represents a 30-s
time interval. Data collection lasted for 2 hours, with a gap owing to
Earth occultation of the line of sight between the NUV and FUV
observations.  An apparent steady decline in flux level is evident,
corresponding to a change of 0.21 mag over the course of 2 hours.
\label{fig:lcurve}}
\end{figure}

The light curve of AM CVn (1650--1700 \AA\ integrated flux) during the
{\it HST} observation is shown in Figure~\ref{fig:lcurve}.  This
wavelength range was observed using both instrument setups and does
not include any strong lines.  There is an apparent, more-or-less
steady decline in flux, which corresponds to a change of 0.21 mag over
the course of 2 hours.

A portion of the apparent decline can possibly be accounted for in
terms of a small inconsistency of calibration between the NUV/G230L
and FUV/G140L observing modes.  Three active galactic nuclei, observed
by one of the authors with STIS using these same modes in the same
back-to-back fashion (albeit with the 52X0.2 aperture rather than the
52X0.1 aperture) are suitable as a test of this, since their UV fluxes
do not vary significantly over the course of a day.  The 1650--1700
\AA\ fluxes differ by about 6--7 percent, in the same sense as
observed for AM CVn.  For AM CVn, the change in the {\em mean}
NUV/G130L 1650--1700 \AA\ flux to the {\em mean} FUV/G140L flux is 15
percent.

Within each observation (NUV or FUV), a declining straight line fit to
the full-band fluxes binned in 30-second intervals gives a
much smaller $\chi^2$ statistic compared with a constant-flux fit, the
slopes being $\sim$0.1 percent per minute in each case.  Taking the
1650--1700 \AA\ fluxes at face value, and considering the NUV and FUV
data as a single series gives an average decline of 0.17 percent per
minute.  Presumptively ``correcting'' the FUV flux relative to the NUV
flux by 7 percent, as suggested by the AGN data, would give an average
rate of decline for the combined data sets of $\sim$0.09 percent per
minute, similar to the individually estimated slopes.
There is, of course, structure in the light curves on top of
straight-line behavior (discussed in \S\ref{variations} below), so
these exercises are only suggestive, not probative.

We have considered whether there is a likely artifactual origin for
the apparent decline in flux, and find nothing plausible.  Thus we
regard the apparent decline as probably real, perhaps modified by a
$\sim7$\% correction to the relative calibration of the FUV and NUV
data.  (We do not apply any such correction further in this paper,
however.) In \S\ref{sect:uvflux}, we discuss other published
observations of AM CVn, as they bear on possible similar flux
variations in UV or visible light.


\subsection{The Mean Ultraviolet Spectrum}\label{sect:meanspec}

\begin{figure}
\includegraphics[angle=-90,scale=.36]{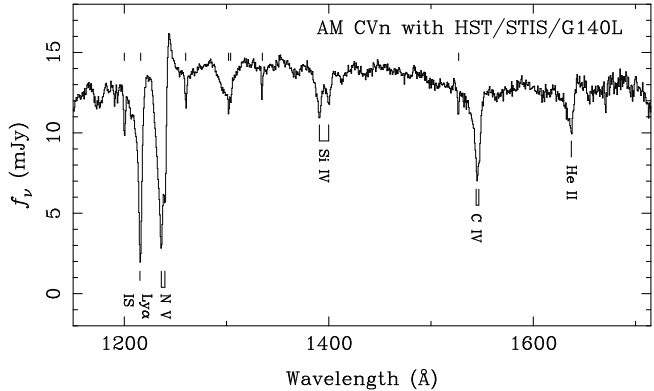}
\caption{Mean short-wavelength
(FUV-MAMA and G140L) spectrum of AM CVn.  The spectrum covers
1150--1715~\AA\ with a resolution of 1.4~\AA.  The strongest
absorption features are marked and labeled below the spectrum; the
tick marks are shifted to align with the troughs of these features.
(See Table~\ref{table:windlines} for laboratory wavelengths; see
Figure 4 for individual profiles).  Sharp, interstellar lines are
indicated with unshifted tick marks above the spectrum (see
Table~\ref{table:islines}).  The ``Lyman-$\alpha$'' feature is
presumably partly interstellar and partly intrinsic, perhaps the
\ion{He}{2} (2-4) transition at 1215.15
\AA.
\label{fig:shortspec}}
\end{figure}

The mean short-wavelength (FUV-MAMA, G140L) spectrum of AM CVn is
shown in Figure~\ref{fig:shortspec}.  The spectrum covers
1150--1715~\AA\ with a resolution of 1.4~\AA\ ($\approx 300~{\rm
km~s^{-1}}$ at mid-range), sampled at $\approx 0.6$~\AA~pixel$^{-1}$.
This spectral region is characterized by a roughly flat continuum (in
$f_\nu$) upon which are superposed strong absorption lines of
\ion{N}{5}, \ion{Si}{4}, \ion{C}{4}, and \ion{He}{2}.  The \ion{N}{5}
and \ion{He}{2} features are flanked by emission on the ``red'' side,
and all the absorption components of the lines are blue-shifted. We
interpret these to be wind lines and discuss their profiles further
below.  These lines are labeled in Figure~\ref{fig:shortspec}.
Table~\ref{table:windlines} summarizes the identifications and
laboratory wavelengths of these features.  An unmarked, broad feature
centered near 1300 \AA\ is likely a blend of numerous (subordinate)
\ion{Si}{3} lines; these and the numerous other weaker features
observable in the spectrum probably arise in the accretion disk and
would be kinematically blended depending on the projected orbital
speed of the gas at their respective radii of formation.

\begin{deluxetable}{lcc}
\tablecolumns{3} 
\tablewidth{3in} 
\tablecaption{Broad UV Absorption Features} 
\tablehead{ 
\colhead{} &
\colhead{$\lambda_{\rm lab}$\tablenotemark{a}} &
\colhead{$\lambda_0$\tablenotemark{b}}  \\
\colhead{Ion}  &
\colhead{(\AA)} &
\colhead{(\AA)}}
\startdata 
\ion{H}{1} Ly$\alpha$, \ion{He}{2}\tablenotemark{c}  & 1215.67, 1215.15 & \nodata \\
\ion{N}{5}              & 1238.82, 1242.80 & 1240.8 \\  
\ion{Si}{4}             & 1393.76, 1402.77 & 1398.3 \\  
\ion{C}{4}              & 1548.20, 1550.77 & 1549.5 \\  
\ion{He}{2}             & 1640.38          & \nodata \\  
\ion{N}{4}              & 1718.55          & \nodata \\  
\enddata 
\tablenotetext{a}{Vacuum (heliocentric) wavelengths are used
throughout.  Observed wavelengths of line cores are often shifted from
the laboratory values, see Figure~\ref{fig:profiles}.}
\tablenotetext{b}{Adopted ``zero-velocity'' wavelength
for profiles shown in Figure~\ref{fig:profiles}, in the case of doublets.}
\tablenotetext{c}{Probable blend of \ion{H}{1} (interstellar) 
and \ion{He}{2} from the AM CVn system, possibly with
a blue-shifted wind component.}
\label{table:windlines}
\end{deluxetable} 



Several sharp interstellar lines are also evident, indicated by tick
marks above the spectrum, and their observed wavelengths, equivalent
widths (EWs), and identifications are given in
Table~\ref{table:islines}.  The average velocity offset of these
measured lines from their adopted laboratory values is $-1~{\rm
km~s^{-1}}$ with a standard deviation of $\sigma(\Delta v) = 36~{\rm
km~s^{-1}}$.



\begin{deluxetable}{lll} 
\tablecolumns{3} 
\tablewidth{3in} 
\tablecaption{Observed Interstellar Lines} 
\tablehead{
\colhead{$\lambda_{\rm obs}$\tablenotemark{a}} &
\colhead{EW\tablenotemark{b}} &
\colhead{}\\
\colhead{(\AA)} &
\colhead{(\AA)} &
\colhead{Identification\tablenotemark{c}}}
\startdata 
1200.29 & 0.47                  & \ion{N}{1}  $\lambda$1199.97 \\
1215.70 & 3.61\tablenotemark{d} & \ion{H}{1} Ly$\alpha$ $\lambda$1215.67 \\
1260.28 & 0.30                  & \ion{Si}{2} $\lambda$1260.42 \\ 
1302.17 & 0.16\tablenotemark{e} & \ion{O}{1}  $\lambda$1302.17 \\ 
1304.14 & 0.10\tablenotemark{e} & \ion{Si}{2} $\lambda$1304.37 \\ 
1334.60 & 0.16                  & \ion{C}{2}  $\lambda$1334.53 \\ 
1526.78 & 0.17                  & \ion{Si}{2} $\lambda$1526.71 \\ 
2343.96 & 0.18                  & \ion{Fe}{2} $\lambda$2344.21 \\ 
2382.51 & 0.27                  & \ion{Fe}{2} $\lambda$2382.77 \\ 
2586.97 & 0.26                  & \ion{Fe}{2} $\lambda$2586.65 \\ 
2599.99 & 0.41                  & \ion{Fe}{2} $\lambda$2600.17 \\ 
2796.55 & 0.34\tablenotemark{f} & \ion{Mg}{2} $\lambda$2796.39 \\ 
2803.38 & 0.42\tablenotemark{f} & \ion{Mg}{2} $\lambda$2803.53 \\
\enddata 
\tablenotetext{a}{Vacuum (heliocentric) wavelengths are used throughout.}
\tablenotetext{b}{The errors on the equivalent widths are $\approx$0.01 \AA, 
unless noted otherwise.}  
\tablenotetext{c}{Only the multiplet components that have lower
$EP = 0$~eV are included; for \ion{N}{1} (UV 1) a multiplet mean wavelength
is quoted; for \ion{Fe}{2} 2382.77 \AA\ (UV 2), the component with the
largest $gf$ value is quoted.  Laboratory data are from Morton
(2003).}  
\tablenotetext{d}{The equivalent width of {Ly}$\alpha$ comes from a
Gaussian fit of the red side of the profile and the core (excluding the
extended blue wing).  If the extended blue wing is included, the EW is
4.26 \AA.}  
\tablenotetext{e}{These two lines are blended and they sit near the
bottom of an absorption feature from AM CVn. Their combined EW has an
error of 0.02 \AA, but the uncertainty in the EW of each individual
line is larger.}
\tablenotetext{f}{These two lines are blended. Their combined EW has an
error of 0.02 \AA, but the uncertainty in the EW of each individual
line is larger.}
\label{table:islines}
\end{deluxetable} 


The \ion{H}{1} ``Lyman-$\alpha$'' feature (rest wavelength 1215.67
\AA) is probably a mixture of an interstellar component and an
intrinsic component, which in this hydrogen-deficient object is likely
the \ion{He}{2} (2-4) transition at 1215.15 \AA.  Like the other
strong absorption lines, this feature is asymmetric, with the blue
absorption wing being broader than the red wing.  This feature is
included in Table~\ref{table:islines}; the tabulated equivalent width
is derived from a Gaussian fit to the core and red side of the line;
if the extended blue wing is included, the EW is 4.26~\AA.  Given the
low resolution offered by the G140L grating, we do not attempt to
decompose this feature or derive a \ion{H}{1} interstellar column
density.

\begin{figure}
\includegraphics[angle=-90,scale=.36]{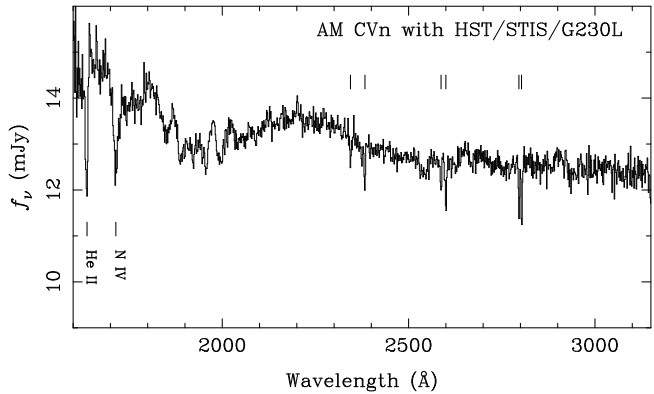}
\caption{Mean long-wavelength
(NUV-MAMA and G230L) spectrum of AM CVn. The spectrum covers
1600--3150~\AA\ with a resolution of 2.8~\AA.  Absorption features
from \ion{He}{2} and \ion{N}{4} are marked and labeled (see
Table~\ref{table:windlines} for laboratory wavelengths; see Figure 4
for the \ion{N}{4} profile).  Sharp, interstellar lines are indicated
with tick marks above the spectrum (see
Table~\ref{table:islines}). Note the expanded vertical scale.
\label{fig:longspec}}
\end{figure}

The mean long-wavelength (NUV-MAMA, G230L) spectrum of AM CVn is shown
in Figure~\ref{fig:longspec}. The spectrum covers 1600--3150~\AA\ with
a resolution of 2.8~\AA\ ($\approx 350~{\rm km~s^{-1}}$ at mid-range),
sampled at $\approx 1.6$~\AA~pixel$^{-1}$.  This spectral region is
also characterized by a roughly flat continuum.  Interstellar lines
(see Table~\ref{table:islines}) are marked above the spectrum as
before; the intrinsic lines of \ion{He}{2} and \ion{N}{4} (see
Table~\ref{table:windlines}) are marked and labeled. (The \ion{He}{2}
line sits in the overlap region common to both the G140L and G230L
spectra.)  The local flux minimum in the $\sim$1800--2000~\AA\ region
(Fig.~3) is not of interstellar origin. The pattern of flux variation
here is similar to that seen in the spectra of early B stars (B0.5 --
B3, especially luminosity classes II and I; cf.\ Wu et al.\ 1991).
Important contributors likely include \ion{Al}{3} and \ion{Fe}{3} (see
the atlases by Rountree \& Sonneborn 1993; and Walborn et al.\  1995).

\begin{figure}
\centerline{\includegraphics[scale=.45]{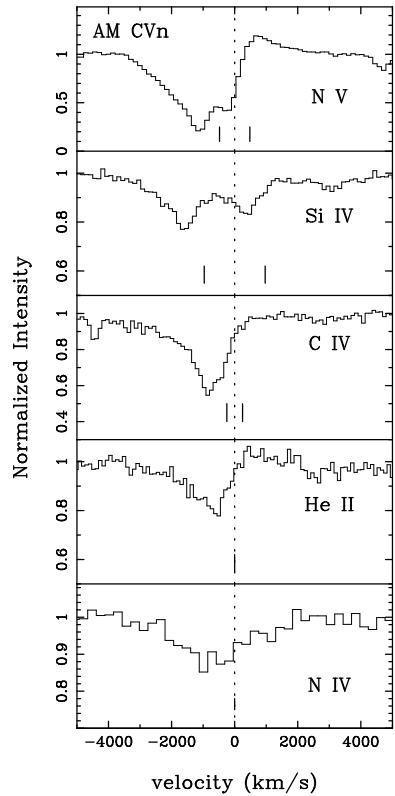}}
\caption{Normalized profiles of some
strong absorption lines are shown on a common velocity scale.  The
velocity zero point is set at the nominal wavelength of a line, or the
average wavelength in the case of doublets; the zero-velocity
wavelengths are given in Table~\ref{table:windlines}.  The tick marks
show the nominal location of the doublet components in the adopted
velocity frame. Normalization is with respect to the local
continuum. The vertical scale varies from panel to panel.
\label{fig:profiles}}
\end{figure}

Mean profiles of the strong absorption lines from \ion{N}{5},
\ion{Si}{4}, \ion{C}{4}, \ion{He}{2}, and \ion{N}{4} are shown in
detail in Figure~\ref{fig:profiles}, on a common velocity scale.  The
velocity zero point in each panel is set at the nominal (rest)
wavelength of the line, or the unweighted average wavelength in the
case of doublets.  (The zero-velocity wavelengths for each feature are
listed in Table~\ref{table:windlines}.)  The tick marks show the
nominal location of each doublet component in the adopted velocity
frame.  All of the observed line cores appear to be blue-shifted by
500--800~km~s$^{-1}$, and \ion{N}{5} and \ion{He}{2} have red-shifted
emission components (P-Cygni profiles), suggesting the presence of an
outflowing wind.  The terminal velocities of the blue-shifted
absorption wings are in the range 2000--3000~km~s$^{-1}$.

\subsection{Short-Term Variations in the Time Domain\label{variations}}

Because the STIS observations of AM CVn were obtained in time-tag
mode, we were able to search for time variations in the spectrum, such
as rapid variations in the line profiles, or short-period periodic or
quasi-periodic variations in the light level.  We discuss first
variations in the line profiles, then variations in brightness.

\begin{figure}
\centerline{\includegraphics[scale=.45]{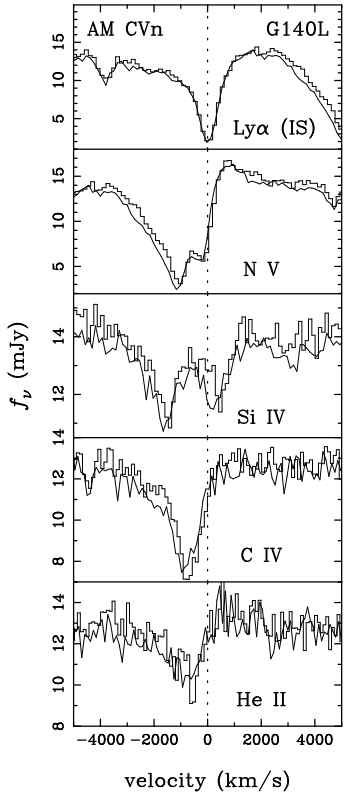}}
\caption{Comparison of the wind line profiles at two epochs.  The
histogram-like lines represent the average spectrum of AM CVn during
480 s (spectra \#10--25 of a sequence of eighty 30-s NUV/G140L binned
spectra), while continuous lines represent the average spectrum during
a later 480 s period (spectra \#30--45).  The gap between the two
intervals is 120 s. The zero-velocity wavelengths used for each
multiplet are as in Table~\ref{table:windlines}.  The (mainly
interstellar) ``Ly$\alpha$'' profile is also shown; its red edge does
not participate in the velocity shifts shown by the wind features.
\label{fig:varyprofiles}}
\end{figure}

We binned the time-tagged observations for both the and FUV/G140L
NUV/G230L data sets, using both 30-second and 3-second bins.  From the
30-second binned data, we prepared trailed spectrograms for each
region. Inspection of the trailed spectrograms (not illustrated) did
not reveal any marked time-dependent behavior. There are, however,
some subtle variations in the profiles of the wind lines, which became
more evident when the mean spectrum was subtracted from the 30-second
binned data.  The pattern of variation comprises an interval of a few
hundred seconds during which both red and blue ``edges'' of the wind
absorption features are located at longer wavelengths than in the mean
spectrum, followed by an interval of similar length in which the
features are shifted to shorter than average wavelengths. The total
displacement corresponds to a few hundred km~s$^{-1}$. The Ly$\alpha$
feature and other interstellar features do not show this variation.
The pattern of red-shift followed by blue-shift is repeated about 900
s later, in the second half of the NUV/G230L observation set.  This
pattern is illustrated in Figure~\ref{fig:varyprofiles}, where spectra
averaged over two 480~s intervals are shown for the (mainly
interstellar) Ly$\alpha$ line and the \ion{N}{5}, \ion{Si}{4},
\ion{C}{4}, and \ion{He}{2} profiles.  The two intervals are separated
by a 120 s gap.  (We did not attempt to optimize the size and spacing
of the intervals chosen for illustration, so 480~s should not be
interpreted as the duration of either the ``red'' or ``blue'' phase.)
The wind from AM CVn thus seems to be mildly unsteady.  Further
interpretation of the wind lines is offered in
\S\ref{sect:lineinterp}.

\begin{figure}
\includegraphics[angle=-90,scale=.385]{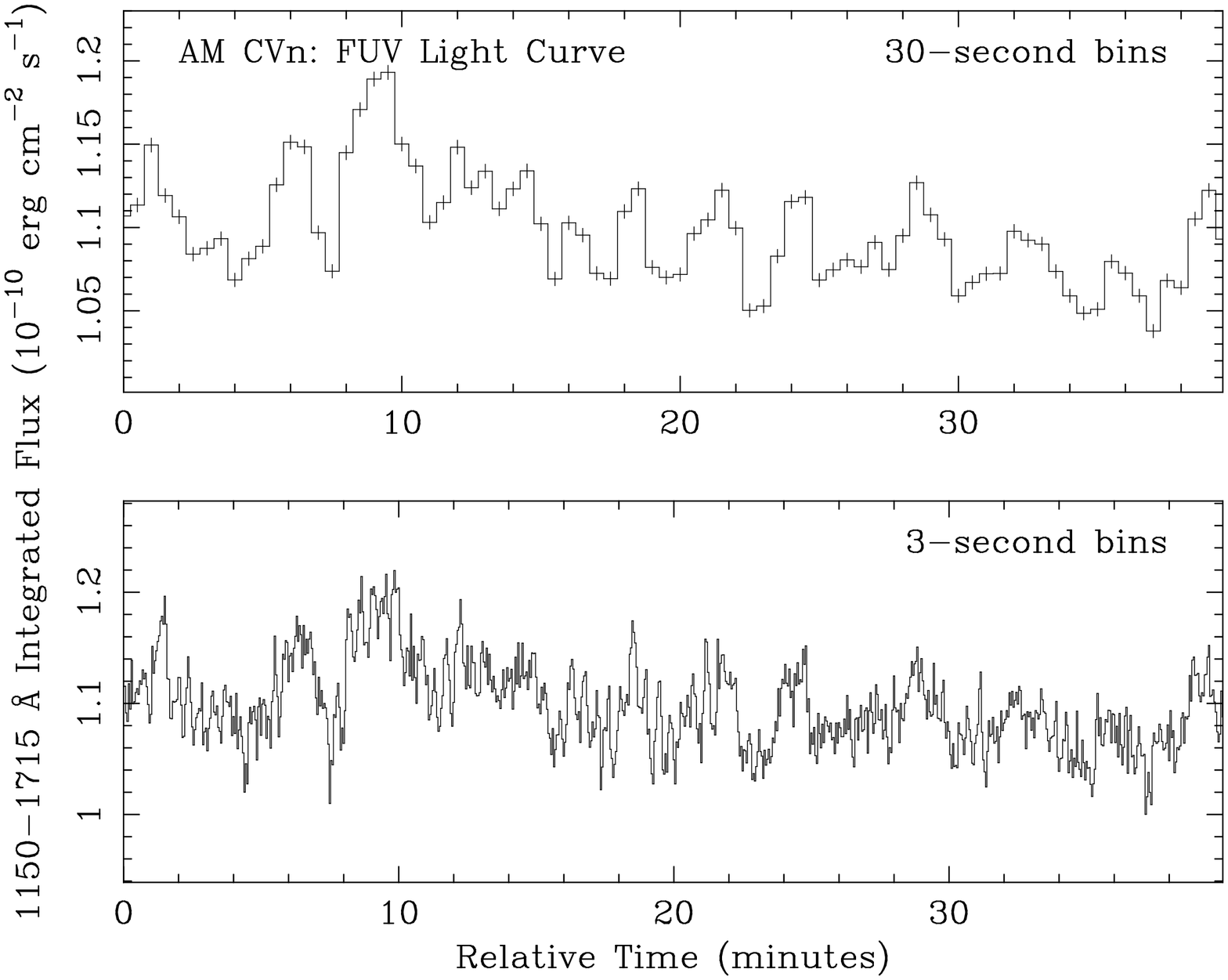}
\caption{Short-wavelength (FUV; 1150--1715~\AA) light curve of AM CVn, 
binned at 30-second and 3-second intervals. 
Error bars are omitted in the latter case for clarity; 
their magnitude is approximately $\pm 2\times 10^{-12}~{\rm
erg~cm^{-2}~s^{-1}}$, i.e., $\pm 1$ minor tick mark on the graph.
\label{fig:short_lc}}
%
\includegraphics[angle=-90,scale=.37]{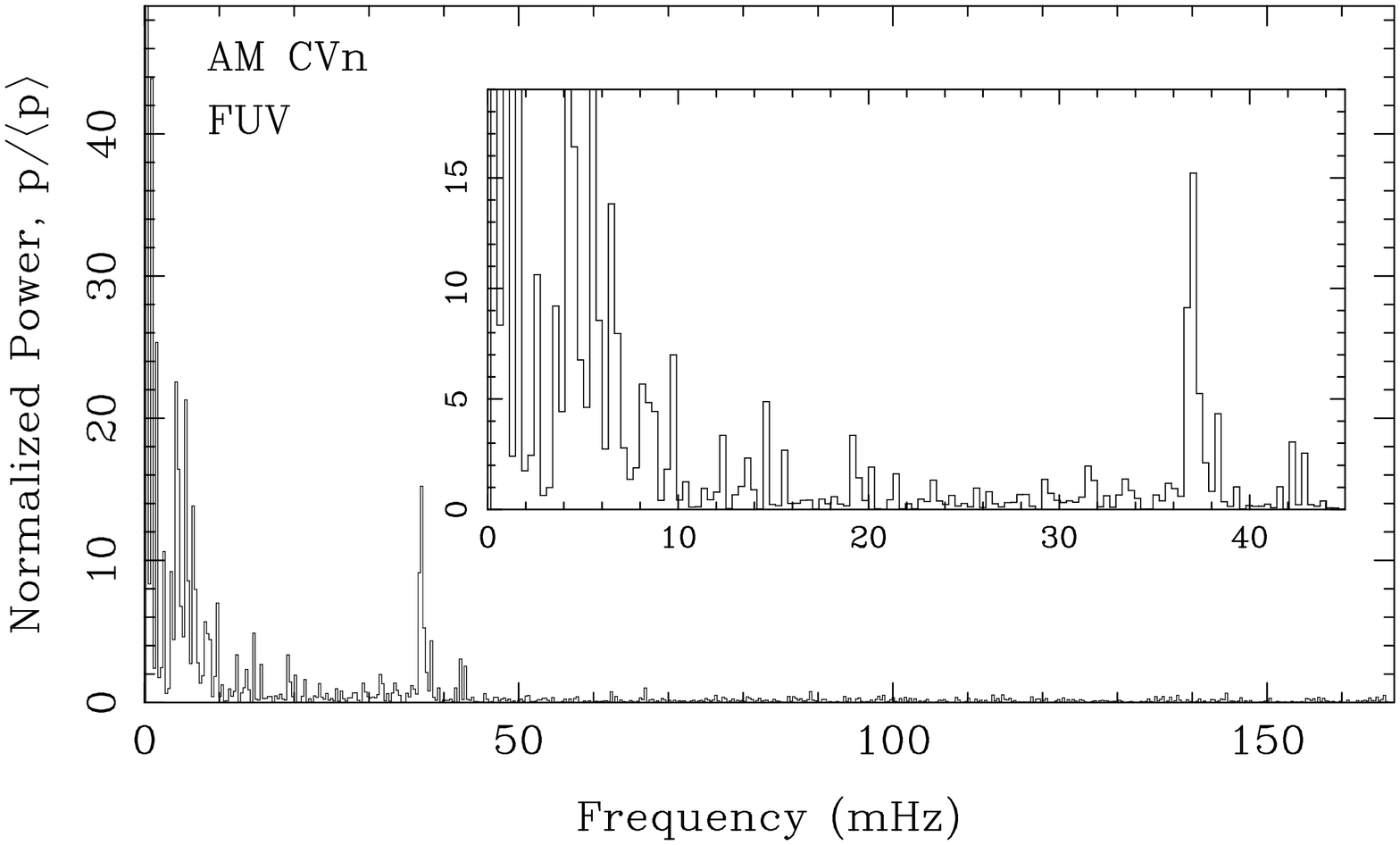}
\caption{Normalized power spectrum for
the short-wavelength (FUV, 1150--1750 \AA) light curve of AM CVn.
Data were binned in 3.0 s intervals, and the Nyquist period is 6
seconds. The power spectrum is sampled at 513 frequencies from 0 mHz to
the Nyquist frequency.  The inset figure shows the low frequency
portion of the spectrum in greater detail.  
In addition to low-frequency noise, a highly
significant, slightly broadened peak ($Q\approx 60$) is seen at
37.0 mHz (${1/\nu}=27.0\pm0.3$~s).  The probability that an isolated
peak of power $p$ will occur by chance is $N \exp({-p/\langle p
\rangle)}$ where $\langle p \rangle$ is the average power
in a spectrum with $N$ bins.
\label{fig:short_power}}
\end{figure}

The 30-second and 3-second binned light curves of AM CVn for the
short-wavelength FUV/G140L data (1150--1715~\AA) are shown in
Figure~\ref{fig:short_lc}.  There is evident structure (flickering) in
the light curve, which is also manifested as low-frequency noise in
the power spectrum.  The power spectrum of the G140L data is shown in
Figure~\ref{fig:short_power}.  The mean level was subtracted
from the 3-second binned light curve, prior to computing the
Fourier transform.  The Nyquist period is 6 seconds, and the power
spectrum is sampled at 513 frequencies.  Each point in the spectrum is
statistically independent of its neighbors, that is, a pure
monochromatic signal would show power in a (centered) single bin.
Normalized power spectra are shown. The significance of a signal is
given in terms of the probability, $N \exp({-p/\langle p \rangle})$,
that an isolated peak of power $p$ could arise by chance in a
white-noise spectrum containing $N$ bins with mean power $\langle p
\rangle$). See Eracleous et al.\ (1991) for further details.

In addition to the low-frequency noise, there is a peak at $f =
37.0$~mHz (${1/\nu}=27.0\pm0.3$~s).  The peak is slightly broadened: a
Gaussian fit shows the FWHM to be 0.61 mHz.  The estimated coherence
of this oscillation is thus $Q \approx 60$, where $Q$ is the ratio of
the peak frequency to the FWHM of the peak. Pure-frequency
(sinusoidal) test signals of various periods injected into the time
series always resulted in power spectrum peaks with $Q \gtrsim 100$.

\begin{figure}
\includegraphics[angle=-90,scale=.41]{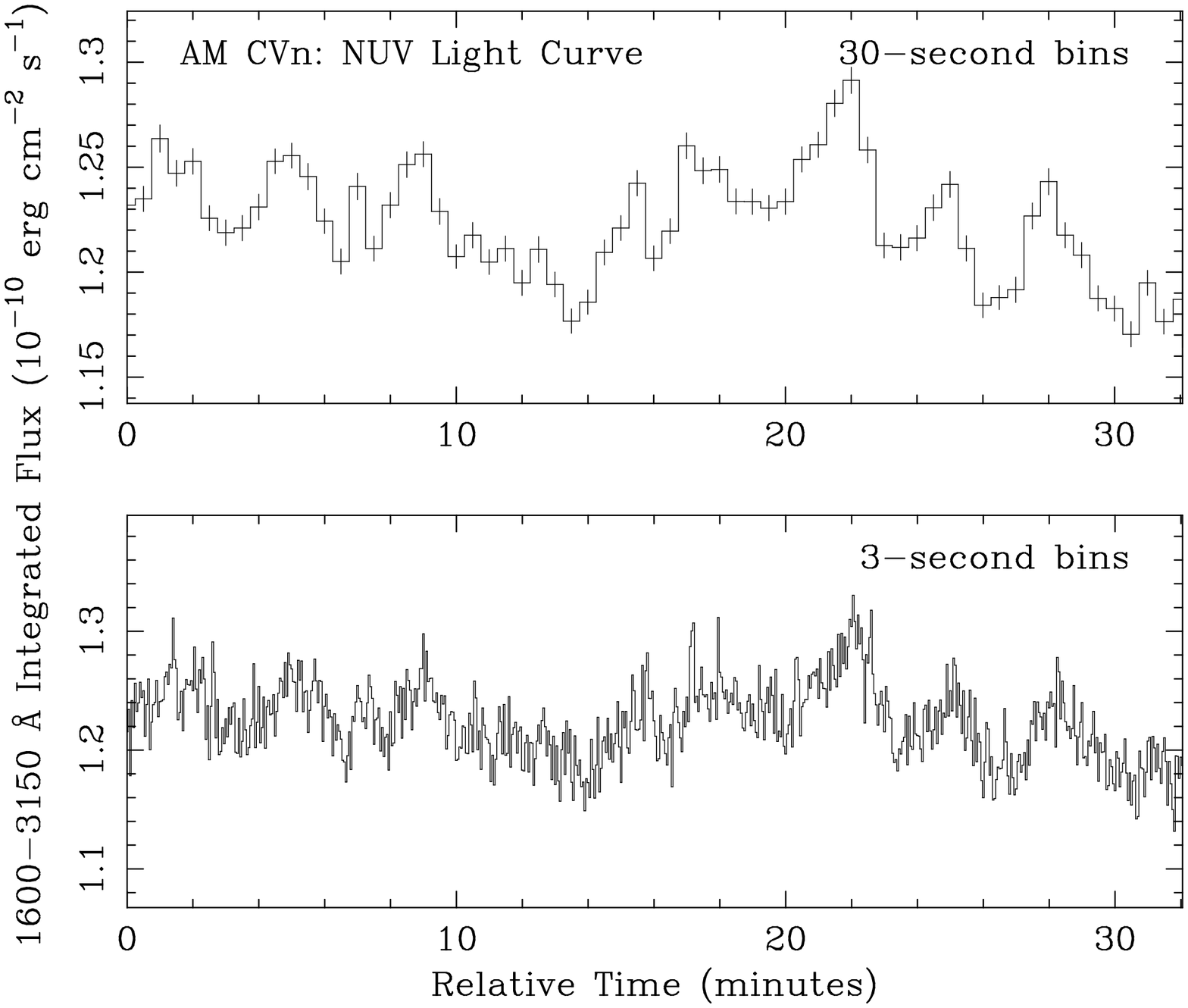}
\caption{Long-wavelength
(NUV; 1600--3150~\AA) light curve of AM CVn, binned at 30-second and
3-second intervals. Error bars are omitted in the latter case for
clarity; they have magnitude approximately $\pm 2\times 10^{-12}~{\rm
erg~cm^{-2}~s^{-1}}$, i.e., $\pm 1$ minor tick mark on the graph.
\label{fig:long_lc}}
%
\includegraphics[angle=-90,scale=.37]{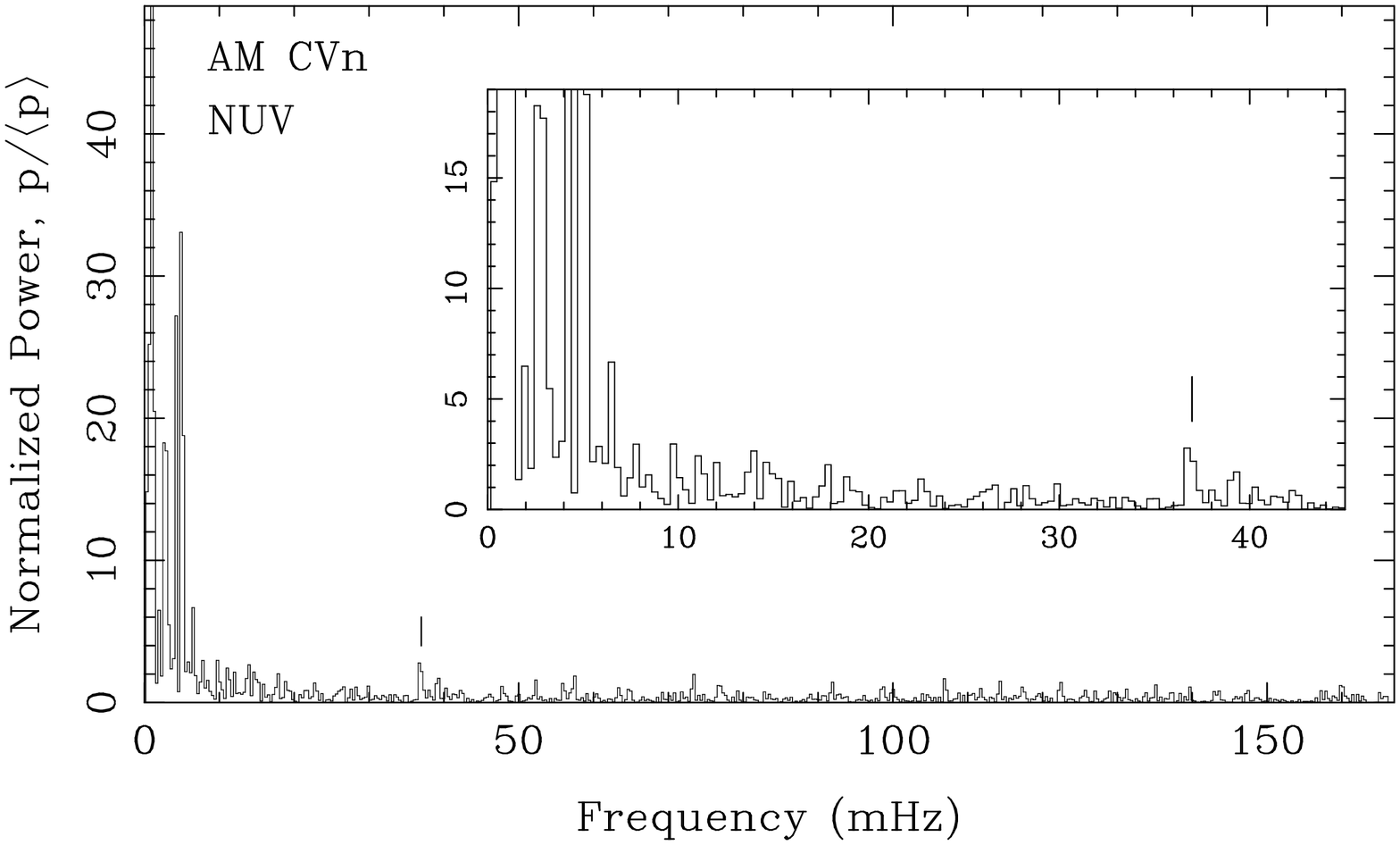}
\caption{Normalized power spectrum for
the long-wavelength (NUV, 1600--3150 \AA) light curve of AM CVn.  Data
were binned in 3.0 s intervals, and the Nyquist period is 6 seconds.
The power spectrum is sampled at 513 frequencies from 0 mHz to
the Nyquist frequency. The tick mark at 37.0 mHz shows the location
of the signal observed in the FUV power spectrum.
\label{fig:long_power}}
\end{figure}

The 30-second and 3-second binned light curves for the long-wavelength
NUV/G230L data (1600--3150~\AA) are shown in
Figure~\ref{fig:long_lc}.  As with the short-wavelength data,
flickering is evident over a range of time scales.  The normalized
power spectrum of the G230L data is shown in
Figure~\ref{fig:long_power}.  As before, the mean level
was subtracted from the light curve, prior to computing
the Fourier transform.  The Nyquist period is 6 seconds, and the power
spectrum is sampled at 513 frequencies.  The power spectrum shows
low-frequency noise, corresponding to the flickering in the light
curve.  The prominent peak at a period of 27~s that is seen in the
short-wavelength power-spectrum is not as obvious here.  While a peak
at that period is discernible, its power is not significantly above
the noise level.

\begin{figure}
\includegraphics[scale=.52]{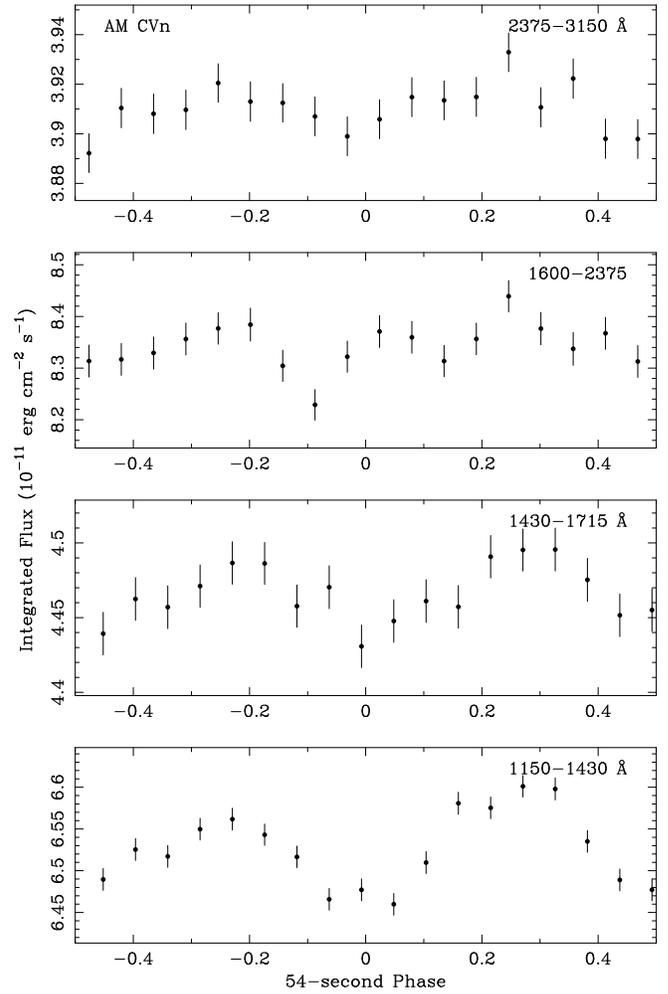}
\caption{Pulse profiles in four different bands,
folded on a 54-second period. The dynamic range is similar in all
frames to facilitate comparison of the relative pulse amplitude
between bands.  The pulse is strongest and the signal-to-noise ratio
highest in the bluest band.  The pulse is somewhat asymmetric,
indicating that the true period may be 54 seconds rather than 27 seconds.
\label{fig:pulseprof}}
\end{figure}

The shape and amplitude of the oscillation vary with wavelength, as
illustrated by the pulse profiles in Figure~\ref{fig:pulseprof}.  The
data are folded on a 54-second pulse period and shown in eighteen
3-second bins.  About forty-four cycles are averaged in the
short-wavelength data, and thirty-five cycles in the long-wavelength
data.  The pulse is strongest and the signal-to-noise ratio highest in
the bluest band (bottom panel), with both quantities decreasing toward
increasing wavelength. The relatively blue spectrum of the pulsed
light explains why the pulses are easily detected in the power
spectrum of the FUV light but are absent in the power spectrum of the
NUV light. The pulse is somewhat asymmetric, indicating that the true
period may be 54 seconds rather than 27 seconds.  Further discussion
of the oscillations is presented in \S\ref{sect:oscns}.

\section{Discussion}\label{sect:discuss}

\subsection{The Ultraviolet Flux Level}\label{sect:uvflux}

We noted in \S\ref{sect:trend} the apparent decline in AM CVn's UV
flux level during our STIS observations. Our tentative conclusion is
that this decline is larger than can easily be accounted for by
instrumental or calibration effects, and should be regarded as real,
especially as it seems to be at least marginally present {\em within}
each of the FUV and NUV observations separately as well as when they
are considered together.

Noting that an excursion of similar amplitude was seen in 1981 March,
but in the visible band and lasting only 500~s (Elsworth et al.\
1982), we investigate whether other brightness changes of similar
rate, duration, or amplitude have been observed in AM CVn before,
either in visible or UV light.

R.\ Kent Honeycutt has kindly communicated to us the results of
synoptic monitoring of AM CVn in the $V$ band over the time period
1990 Nov -- 1996 Aug. These observations were made using the Roboscope
(Honeycutt et al.\ 1994). One to four magnitude measurements were made
per night, with 506 measurements in total spread over 417 nights.  On
nights with multiple observations (usually separated by a few hours)
the intranight variations in brightness are consistent with the
measured errors of observation, which are typically at or below the
one per cent level.  On a few occasions brightness changes of a few
per cent from one night to the next are noted, although they are of
marginal statistical significance. Slow drifts in the average $V$
magnitude of a similar amplitude are also noted over the course of an
observing season.  No ``outbursts'' are seen, nor any rapid and
persistent changes in brightness such as those seen in the STIS data
of Figure~\ref{fig:lcurve}.  The median magnitude is $V=14.11$.

Skillman et al.\ (1999) also report from high-speed photometry that
the visual brightness of AM CVn is quite robust, always between
$V=14.10$ and 14.20 and never varying more than 0.07 mag during a
night.

Taking the various calibrated fluxes at face value, we find the UV
flux from AM CVn to vary somewhat over {\em long} intervals, as shown
by the mean UV spectrum of AM CVn obtained in 1995 with the GHRS in
program 6085. The s.e.d.\ is flat ($f_\nu \approx$~constant) over the
wavelength range 1260-1560 \AA, with a mean continuum flux level of
about 16.5 mJy, or about 15\% higher than our nominal FUV/G140L mean
flux level. 

We have also compared the UV brightness of AM CVn found in our STIS
observations with the flux level measured by IUE at several epochs
between 1978 and 1988.  (See Boggess et al.\ 1978 for a description of
IUE instrumentation.)  We considered only IUE observations made in the
LOW dispersion mode using the LARGE aperture.  We retrieved NEWSIPS
(Final Archive) fluxes in ASCII form from the Browse facility at
MAST\footnote{Multimission Archive at the Space Telescope Science
Institute,\\ http://archive.stsci.edu/iue/} and computed average flux
densities for each observation, in 60 \AA\ (100 \AA) intervals
centered at 1450 \AA\ or 2700 \AA, depending on the camera used.  At
1450 \AA, the IUE fluxes range between 2.1 and $2.4 \times
10^{-13}~{\rm erg~cm^{-2}~s^{-1}}$~\AA$^{-1}$, and the corresponding
STIS measurement is 2.0 in the same units.  At 2700 \AA, the IUE
fluxes lie between 5.5 and $6.0 \times 10^{-14}~{\rm
erg~cm^{-2}~s^{-1}}$~\AA$^{-1}$, and the STIS measurement is 5.1 in
the same units.  Thus the STIS observations show AM CVn to be somewhat
less bright in 2002 February than was recorded by IUE, although by
only $\sim$10\%, comparable to the range of variation observed with
IUE.  Massa \& Fitzpatrick (2000) point out some concerns about the
NEWSIPS absolute calibration of IUE fluxes, at the 10\% level, so our
comparison of IUE and STIS fluxes is preliminary only.

Ramsay et al.\ (2005) present near-UV observations of AM CVn with the
Optical Monitor aboard the XMM-Newton observatory, spanning an
interval of about 12000 s.  The effective wavelength was 2910~\AA\ 
(UVW1 filter, range 2400--3400 \AA). Excepting a few outliers, count
rates in 120 s bins do not show deviations from the mean rate that are
larger than about 3\%. This rate is not easily converted to absolute
flux units for comparison with other instruments.

In physical terms, it is not difficult to imagine that variations in
UV brightness at the $\sim$10\% level might occur, driven by
variations in accretion rate through the disk and onto the central
white dwarf.  One might expect larger variations at UV wavelengths
than in visible light, and indeed the higher amplitude at shorter
wavelengths of the 27-s oscillations seen in the STIS data are an
example (\S\ref{variations}).  Other cataclysmic variables (CVs) in a
persistent high-luminosity state have also been observed to vary in
the UV.  Hartley et al.\ (2002) document $\sim$30\% variations in the
continuum level of V3885 Sgr and less certain variations at the
factor-of-two level in IX Vel, between observations with STIS spaced
weeks or months apart.

To summarize, AM CVn's brightness in the UV {\em sometimes} appears to
be more strongly variable than in the visible on time scales of hours
or longer, although the data are too sparse to allow full
characterization of this variation, and some calibration uncertainties
persist at the few percent level.  Without further time-series
observations in the UV, we cannot say whether or not this behavior
(hours-long trends) happens frequently.

\subsection{The oscillations}\label{sect:oscns}

Our power spectrum analysis of the UV light curve of AM CVn
(\S\ref{variations}) revealed a slightly broadened peak at 37 mHz,
corresponding to an oscillation period of 27.0 s.  Folding the light
curve shows a waveform that is approximately sinusoidal, although
alternating maxima have slightly different heights suggesting the
underlying period may be $\sim$54 s. This is only a 2--$\sigma$
effect, however, and we discuss the oscillation in terms of a 27 s
period.  The amplitude (half of the peak-to-peak variation) is
$\sim$1\% in the shorter wavelengths, diminishing to about half this
in the longest observed wavelengths (Figure~\ref{fig:pulseprof}).

Patterson et al.\ (1979) reported a ``transient coherent or
quasi-coherent periodicity at 26.3 s'' in visible-light time series
photometry of AM CVn. Patterson et al.\ (1992) showed the power
spectrum of this oscillation. The amplitude was $\sim$0.01 mag,
uncertainly measured because of the poor coherence of the signal; they
referred to the signal as a ``quasi-periodic oscillation'' and revised
the mean period to 26.2 s.  In a recharacterization, Skillman et al.\
(1999) stated that ``the period and coherence of this signal are quite
plausible for `dwarf nova oscillations'.''  We adopt the point of view
that the signal seen by Patterson et al.\ (1979) and by us at
different times and in different wavebands arises from the same cause.

Are these variations ``dwarf nova oscillations'' (DNO), or are they
``quasi-periodic oscillations'' (QPO)?  According to the review by
Warner (2004), DNOs are moderately coherent ($Q_W \equiv |dP/dt|^{-1}
\gtrsim 10^3$), nearly-sinusoidal signals with periods typically in
the range 8 to 40~s.  QPOs are less coherent, with periods typically
$\sim$10 times longer than the corresponding DNOs for a CV.  DNOs and
QPOs may appear simultaneously or not.  DNOs are characteristically
observed during the outbursts of dwarf novae, but are also seen in
some luminous ``novalike'' CVs.  In an outburst, the period of the DNO
varies with time, inversely related to the luminosity of the dwarf
nova.  In addition to this luminosity-linked variation, the DNO period
may show sudden small jumps, best studied by means of an amplitude and
phase analysis, in which short segments of the light curve are fitted
directly in a sliding window using a sinusoidal model for the
variation.  Given the similarity of the DNO period to the Keplerian
period in the inner disk, DNOs are interpreted as having their origin
at or near the surface of the accreting white dwarf or in the inner
disk. Warner (1995) argues that QPOs, with their longest periods
corresponding to the orbital timescale in the outer disk, may be
linked to oscillations of the disk, with a mixture of periods
corresponding to different radial zones.

If the 27-s (37 mHz) signal from AM CVn is a DNO, we might look for a
QPO near 4--5 mHz. The power spectrum shows a ``grassy'' structure at
these frequencies, which may indicate the presence of a QPO or may
simply be part of poorly-defined ``1/f noise'', characteristic of
low-level ``flickering''.  More data would be required to develop a
well-averaged power spectrum that would show whether a broad QPO
``bump'' lies on top of a 1/f ``continuum''.

We performed a simple analysis of amplitude and phase for the FUV
light curve of AM CVn, using a moving window 270 s (10 cycles) long.
We fitted a model consisting of a linear baseline and a pure sinusoid
with fixed period.  The form of the baseline is $A + Bt$, and $A$ and
$B$ vary as the window is shifted along, to take out some of the
low-frequency wandering in the light curve. We note well-defined
``DNO-like'' phase drifts of duration several hundred seconds,
indicating a period that is changing around the mean period, as in
Warner (2004; his Fig.~2).  We also note that the amplitude of the
fitted sinusoid is typically higher in the first half of the time
series than later (in our Figure~\ref{fig:short_lc}, there is high
amplitude around Relative Time = 16 minutes, and low amplitude around
26 minutes, as extreme examples).  The width (FWHM) of the spike in
the power spectrum partly arises from this amplitude modulation
(effectively, only half of the time series is useful in establishing
the frequency of the signal), and partly arises from the small drifts
of period during the half of the observation where the oscillation is
strong.

Despite the difference in chemical composition of AM CVn's disk
compared with the usual hydrogen-rich CV, the temperature of the inner
disk is likely similar to a dwarf nova in outburst.  It would
therefore not be surprising to observe DNOs from AM CVn in the same
period range as normally observed. The amplitude of a DNO in the UV
would be high relative to that in visible light, since the UV light
from the inner disk is not diluted by light from the outer disk.  We
therefore adopt the interpretation of Skillman et al.\ (1999) in
characterizing the 27-s oscillation as a DNO.  We note, however, that
the defining test of whether the period changes with system luminosity
cannot be carried out, since AM CVn does not undergo ``dwarf nova''
outbursts.  Labeling the oscillation as a DNO does not tell us the
cause of the oscillation, which seems to be still very open to
debate. Different points of view as to the physical mechanisms
involved are represented by Warner \& Woudt (2006), Mauche (2002), and
Piro \& Bildsten (2005).

Since the ``outer disk'' in AM CVn is less extensive than in
longer-period CVs, one is observing the inner disk with relatively low
dilution by lower temperature gas, even in visible light.  It might
therefore be regarded as slightly puzzling that the $\sim$26 s
oscillation has not been reported more frequently.  But the source is
relatively faint, and attention historically has been focused on the
``orbital/superhump'' modulation at 1029/1051 s, both factors tending
to the use of integration times that are long compared with 26 s.  In
the ultraviolet, the same is true.  The analysis by Solheim et al.\
(1997) of time-resolved UV spectra, obtained with the GHRS, made use
of data binned in $\sim50$~s intervals (with gaps).  It thus is not
sensitive to the frequency range that includes the signal we report
here.  Ramsay et al.\ (2005) report a power spectrum of the integrated
UV light from AM CVn, from their observations using the XMM-Newton
Optical Monitor. It shows significant power near periods of 996, 529,
and 332 s.  Their analysis does not extend to periods as short as 27
or 54 s, however.

\subsection{The wind lines}\label{sect:lineinterp}

The blue-shifted broad absorption features in the UV spectrum of AM
CVn (Table 1 and Figures 4 \& 5), with red-shifted emission apparent
only in \ion{N}{5} and perhaps \ion{He}{2}, are typical of the
wind-formed lines seen in the spectra of all luminous CVs seen at low
or moderate inclination.  In AM CVn, the deepest lines are \ion{N}{5}
and \ion{C}{4}. The terminal velocity of the absorption trough is near
$-3000~{\rm km~s^{-1}}$, most cleanly seen in \ion{N}{5}.

Time-resolved UV spectra of AM CVn, obtained in 1995 with the GHRS,
have apparently not been published in detail, although a mean spectrum
and some time-series analysis are reported in Solheim et al.\ (1997).
A superficial comparison of the mean GHRS and STIS
spectra in the region of overlap shows that they are very similar,
apart from the difference in overall brightness level mentioned
earlier.

Proga (2005) reviews the modeling of CV winds, while Froning (2005)
reviews the observations and highlights some of the difficulties that
models have in reproducing the diversity of behavior that is seen in
the lines.  Kinematic models of CV wind lines (e.g., Knigge \& Drew
1996) explain them as being formed mainly by scattering in a
non-spherically symmetric, or ``bi-conical'' outflow.  The winds lines
are stronger in higher luminosity CVs, suggesting that they are driven
by radiation pressure. Two-dimensional, time-dependent hydrodynamic
models of radiation-driven winds indeed produce a slow, dense
``equatorial'' wind bounded on the pole-ward side by a faster more
tenuous flow (Proga, Stone \& Drew 1998, 1999).  The winds are
predicted to be stronger for a higher luminosity system. The terminal
velocities are a few times the escape velocity from the white
dwarf. If the disk luminosity is higher than that of the central white
dwarf, the outflow tends to become unstable and clumpy.  Some
qualitative success has been achieved at matching predicted line
profiles with observations, although problems remain, and a unique
matching of a model to a data set is elusive (e.g., Proga et al.\
2002; Proga 2003; Long \& Knigge 2002; Froning 2005).

We have made a qualitative comparison of the mean profiles from Figure
4 with illustrative computed profiles presented in Proga et al.\
(2002), Long \& Knigge (2002), and Proga (2003). We concentrate mainly
on the \ion{C}{4} profile.  AM CVn is not an eclipsing binary system,
so an intermediate inclination angle is inferred.  Consistent with
this, the overall observed shape in \ion{C}{4} (terminal velocity,
concavity, and location of maximum depth close to zero velocity)
matches reasonably well to model profiles calculated for intermediate
inclinations ($i=30^\circ$ or $55^\circ$).  The fixed parameters of
these models (white dwarf mass and radius, disk radius, etc.) are not
tuned to AM CVn, and the sampling of the variable parameters
(inclination, wind mass loss rate) is too coarse to allow more
quantitative conclusions to be drawn, but the explanation of the
observed line profiles in terms of a bi-conical wind seems to be
secure.

It is reassuring that CV wind theory seems to be applicable in the
case of hydrogen-deficient, ``ultracompact'' binaries such as AM
CVn. Nor is this a surprising result, if the wind originates in the
inner disk and is viewed against continuum light from the inner disk.
The outer disk is ``missing'', but seems to have little effect on the
shaping of the {\em absorption} part of the line profile, and the UV
opacities in the UV-forming inner disk are affected at only the
factor-of-two level by the absence of hydrogen.  The shape of the
ionizing continuum from a hydrogen-deficient disk is expected to be
different, however, so the interpretation of the strength of a wind
line profile in terms of a wind mass loss rate may be expected to
differ from the hydrogen-dominated case. (The strength of the line
depends on the population of the ion, hence on the ionization state of
the gas as well as the overall density of the wind).

The shifting positions of the resonance line absorption features shown
in Figure~\ref{fig:varyprofiles} is a small but real effect, likely
showing that conditions in the outflow are not steady.  In AM CVn, the
profiles as a whole seem to shift back and forth, maintaining their
shapes. Certain other luminous CVs show much more ``wild'' behavior in
the wind lines, e.g., V603 Aql (Prinja et al. 2000a) and BZ Cam
(Prinja et al.\ 2000b), in which the shapes of the lines and their
equivalent widths change drastically, and individual ``blobs'' or
clumps may be traced as their velocities change with time.  The theory
of radiation-driven winds suggests that the outflow becomes unstable
and clumpy if the disk luminosity (vertically directed) dominates over
the radially-directed luminosity from the central star (Proga et al.\
1998).  There is thus some hope that this aspect of disk wind theory
can be confronted by the AM CVn observations, where the instability
seems to be marginal, if the disk and white dwarf luminosities can be
determined.  In the case of RW Sex, on the other hand, the time
variations in the profiles of far-UV wind lines seem to be related to
orbital modulation, and any fluctuations of density or speed in the
wind may occur on scales small enough that they do not appear with
adequate contrast in the observed line profiles (Prinja et al.\ 2003).
We noted in \S\ref{variations} a rough time scale for the observed
line variations of $\sim$900 s, comparable to the orbital period of AM
CVn (1029 s), but only further observations could establish whether
there is a strict correspondence. (Interest is added, owing to the
fact that the secondary star in AM CVn, although small, is much closer
to the source of the wind and thus might present a means of modulating
the outflow on the orbital period.)  At present, our only firm
conclusion is that there are variations present in the wind line
profiles, with the cause yet to be determined.

\subsection{The observed spectral energy distribution}\label{sect:obsdsed}

\begin{figure}
\includegraphics[scale=.48]{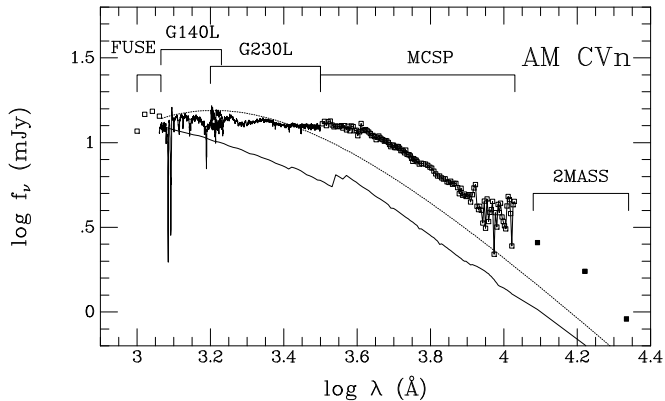}
\caption{Spectral energy distribution of AM CVn, assembled from
various sources at various epochs (see \S\ref{sect:obsdsed}).  No
correction for interstellar extinction has been made.  Also shows are
example calculations of the s.e.d.\ from a (non-irradiated)
steady-state accretion disk specified by parameters suggested by
Roelofs et al.\ (2006); see \S\ref{sect:obsdsed} for details.  The
upper (dotted) curve shows a model that uses blackbody spectra to
represent the local emission from the disk surface; the lower (solid)
line uses interpolated spectra from pure helium model atmospheres
instead, neglecting limb darkening. No other sources of light are
included. 
\label{fig:sed}}
\end{figure}

We show in Figure~\ref{fig:sed} the observed s.e.d.\ of AM CVn,
assembled from several different sources.  Data at the shortest
wavelengths come from FUSE observations (epoch 2004 Dec 26), as read
by eye from ``preview'' spectra provided at the MAST website{\footnote
{\tt http://archive.stsci.edu/fuse/}}; the points represent estimated
continuum fluxes at 1000, 1050, 1100, and 1150 \AA.  Next are the STIS
data reported earlier in this paper (``G140L'' and ``G230L'').
Optical and near-infrared data (``MCSP'') are from the Multichannel
Spectrophotometer formerly in use at the Hale 5m telescope.  These
data are described in Oke \& Wade (1982), and can be made available in
machine-readable form on request; we suggest that these
quasi-monochromatic flux density measurements be used in place of
broadband $UBVRI$ magnitudes when modeling the s.e.d.  Finally the
infrared (IR) broadband fluxes from 2MASS (Skrutskie et al.\ 2006) are
given, using the absolute calibration and zero-point offsets from
Cohen et al.\ (2003). We summarize these data here: isophotal
wavelengths of the 2MASS $J$, $H$, and $K_s$ bands are 1.235, 1.662,
and 2.159 microns respectively; the corresponding isophotal
monochromatic flux densities are 2.57, 1.74, and 0.91 mJy with random
errors of 3\%, 4.5\%, and 6.6\%. No correction for interstellar
extinction has been made in Figure~\ref{fig:sed}.

The GHRS fluxes (1260--1540 \AA) from 1995 lie about 15\% (0.06 dex)
higher than the STIS/G140L fluxes shown in the figure.  The GHRS
fluxes would more smoothly connect the STIS/G230L and FUSE data sets.
This points up a general problem with modeling the s.e.d.\ of AM CVn,
namely what fluxes are to be modeled.  In view of the mixed nature of
the data --- some broadband, some quasi-monochromatic; obtained at
different epochs with evidence of variability; etc. --- an attempt to
find an ``optimum'' fit to the s.e.d.\ should give careful
consideration to the figure of merit that is used to determine
goodness of fit, including, e.g., how the various parts of the
spectrum are weighted.

The spectrum of AM CVn can be modeled to infer $dM/dt$ and perhaps
other quantities such as $M_1$, $M_2$ (the mass of the donor star),
the distance $d$, and the inclination $i$ of the disk orbital plane to
the line of sight.  An accurate knowledge of $dM/dt$ is needed to (1)
confirm that the disk in AM CVn satisfies the thermal stability
criterion for disks in a permanent ``high'' state (Smak 1983; Tsugawa
\& Osaki 1997); (2) assess the thermal and evolutionary properties of
the donor star (e.g., Deloye et al.\ 2005); and (3) properly partition
the observed flux between the disk and the other contributors to the
radiation.  Item (3) is necessary for deriving abundances from
modeling of the absorption line spectrum in the UV, where some of the
light may come from the central white dwarf and some from the disk.

The library of local disk spectra that is used to construct a model
s.e.d.\ of the entire disk is of paramount importance.  Blackbody
spectra, for example, differ greatly from the s.e.d.s of helium
atmospheres.  We illustrate in Figure~\ref{fig:sed} the difference
between a blackbody disk (BBD) and a helium-atmo\-sphere disk (HeD),
computed with the same model parameters.  Our model disk is the usual
steady-state, geometrically thin, optically thick, non-irradiated
accretion disk. We choose $M_1= 0.69~{\rm M_\sun}$, $R_1 = R_{\rm wd}
= 7.8 \times 10^8$~cm, $R_d = 7.6 \times 10^9$~cm, $dM/dt = 6.7 \times
10^{-9}~{\rm M_\sun~yr^{-1}}$, $d= 600$~pc, and $i= 43\degr$, similar
to the model for AM CVn proposed by Roelofs et al.\ (2006).  Here $R_1
= R_{\rm wd}$ is the inner radius of the disk, taken to be the radius
of the white dwarf; and $R_d$ is the outer radius of the disk.  For
this illustration we have used angle-averaged fluxes from Wesemael
(1981) and D.\ Koester, private communication. No sources of light
other than the disk are included.  The UV-optical-IR model fluxes
from the HeD are a factor of $\sim1.6$ lower than those from the BBD,
owing to different bolometric corrections in the two cases (the
``missing'' flux is emitted in the extreme ultraviolet). With the chosen
parameters, the overall fit of the HeD model is worse than that of the
BBD.  By increasing $dM/dt$, the HeD model can be made to better match
the normalization of the observed s.e.d.

Independent constraints on $d$, $M_1$, $R_1$, $M_2$, and $i$, if they
are trustworthy, are very important for reducing the number of
correlated parameters or the degree of correlation in a model of the
accretion disk.  In the AM CVn system, two constraints are especially
significant for inferring $dM/dt$ from the s.e.d. The first is the
system's mass ratio $q$, for which estimates range from $q=0.087$
(Nelemans et al.\ 2001) to $q=0.22$ (Pearson 2003).  Roelofs et al.\
(2006) have put forward a spectroscopically determined $q=0.18 \pm
0.01$.  Once $q$ is determined, $M_1$ and $M_2$ cannot be varied
independently.  Varying $M_1$ affects both temperatures and orbital
speeds (line smearing) in the disk, as well as affecting the
fundamental solid angle $(R_1/d)^2$; hence $dM/dt$ as inferred from
observation is also affected. Meanwhile $dM/dt$ is tied to $M_2$
through the theory of mass transfer driven by angular momentum loss.

The second constraint is the distance.  The often-used value $d =
235$~pc is based on an unpublished parallax, $\pi_{\rm abs} = 4.25 \pm
0.43$~mas from the USNO parallax team (C.\ Dahn, private communication
2003 \& 2006; 62 observations over 8.1~y; seven stars in the
astrometric reference frame).  Recently a new parallax determination
using the Fine Guidance Sensor aboard the HST, $\pi = 1.65 \pm
0.30$~mas ($d \approx 606$~pc), has been published (Roelofs et al.\
2007).  This study has a duration of 1.6~y and relies on three
astrometric reference stars, all to the south of AM CVn.  In
Figure~\ref{fig:sed}, reducing $d$ from 600 pc to a smaller value
would raise the predicted fluxes of the model for a given $dM/dt$.  A
substantially smaller distance such as $d = 235$~pc would therefore
mandate a reduction in the model $dM/dt$ in order to match the
observed flux level, leading also to a cooler and ``redder'' model
disk; in the case of the HeD model, the model would better match the
observed relative s.e.d.  The formal difference between the two
parallax measurements is statistically very significant, and $d^2$
differs by a factor of $\approx$7.  Choosing to apply one of the
advertised distance estimates over the other, as a modeling
constraint, has a huge impact on whether any particular model for AM
CVn can successfully account for both the s.e.d.\ and spectroscopic
observations of that system.  We are not in a position to resolve this
discrepancy.

The strong linkages among $q$, $d$, $dM/dt$, and other system
parameters are illustrated in the analysis of several AM CVn-type
systems by Roelofs et al.\ (2007).


\section{Summary}\label{sect:summary}

We have presented new, time-resolved STIS observations of AM CVn,
which show a UV spectrum that is approximately flat in $f_\nu$.  The
absorption profiles of \ion{N}{5} $\lambda$1240, \ion{Si}{4}
$\lambda$1398, \ion{C}{4} $\lambda$1549, \ion{He}{2} $\lambda$1640,
and \ion{N}{4} $\lambda$1718 are asymmetric and blue-shifted,
evidencing a wind that is partly occulted by the accretion disk.
There is also weak emission at \ion{N}{5} and \ion{He}{2}.  These
features are consistent with profiles predicted for scattering in a
bi-conical wind from an accretion disk, viewed at intermediate
inclination. The profiles of these wind lines vary mildly with time,
showing shifts in the observed wavelengths of both red and blue
absorption edges.  Sharp (interstellar) absorption lines are also
seen.  The \ion{H}{1} Lyman-$\alpha$ feature is presumably
interstellar, but may be blended with the \ion{He}{2} (2-4) transition
arising in AM CVn itself. Numerous weaker spectral features of various
widths are found, probably arising in the accretion disk and
kinematically blended.

The UV light curve of AM CVn from the STIS observations shows an
apparent, relatively steady decline by $\sim$20\% over the span of the
observations.  Only a portion of the nominal decline can be attributed
to possible calibration uncertainties. We summarize data that suggest
AM CVn's brightness varies by a larger amount in the UV than in the
optical.  There are also short-term ``white light'' variations,
including a 27-s DNO that is stronger at shorter wavelengths.  The
true period of the DNO may be 54 s.

We have assembled the UV-visible-IR s.e.d.\ of AM CVn by combining the
STIS observations with data from FUSE, the Palomar MCSP instrument,
and 2MASS.  Successful models of the accretion process in AM CVn,
accounting for the shape and normalization of the entire observed
s.e.d., may give a robust estimate of the mass accretion rate $dM/dt$
and other parameters of the system.  The mass accretion rate is of
great interest to understanding the origin and subsequent evolution of
HeCVs.  Inferences about $dM/dt$ depend strongly on the local
radiative properties of the gas that makes up the accretion disk.  We
have illustrated this by explicit computation of the s.e.d.\ from a
blackbody disk and a Helium-atmosphere disk, using the same example
specification of $M_1$, $R_1$, $R_d$, $dM/dt$, $d$, and $i$ in both
cases. The results are quite different.  Other key factors in
inferring $dM/dt$ include what is assumed about the mass of the
accreting star (strongly influenced by the adopted mass ratio $q$) and
the distance to the system.

\acknowledgments

We are grateful to R.\ K.\ Honeycutt for communicating to us the
results of Roboscope synoptic observations of AM CVn, to M.\ Rogers
for assistance in the review of literature, and to C.\ Dahn for some
correspondence about the USNO parallax result. D.\ Koester kindly
provided some model spectra for pure helium atmospheres. We benefitted
from helpful comments by an anonymous referee.  M.E. acknowledges the
hospitality of the Astrophysics Department at the American Museum of
Natural History.

This research has made use of the SIMBAD database, operated at CDS,
Strasbourg, France.  This publication makes use of data products from
the Two Micron All Sky Survey, which is a joint project of the
University of Massachusetts and the Infrared Processing and Analysis
Center/California Institute of Technology, funded by the National
Aeronautics and Space Administration and the National Science
Foundation.  Some of the data presented in this paper were obtained
from the Multimission Archive at the Space Telescope Science Institute
(MAST). STScI is operated by the Association of Universities for
Research in Astronomy, Inc., under NASA contract NAS5-26555. Support
for MAST for non-HST data is provided by the NASA Office of Space
Science via grant NAG5-7584 and by other grants and contracts.
Support for HST GO program \#8159 was provided by NASA through a grant
from the Space Telescope Science Institute, which is operated by the
Association of Universities for Research in Astronomy, Inc., under
NASA contract NAS 5-26555.



\end{document}